\documentclass[12pt]{article}
\usepackage{epsfig}
\font\mybb=msbm10 at 12pt
\font\mybblarge=msbm10 at 16pt
\font\mybbslightlylarge=msbm10 at 14pt
\font\mybbsub=msbm10 at 8pt
\font\mybbsmall=msbm10 at 10pt

\def\bb#1{\hbox{\mybb#1}}
\def\bblarge#1{\hbox{\mybblarge#1}}
\def\bbslightlylarge#1{\hbox{\mybbslightlylarge#1}}
\def\bbsub#1{\hbox{\mybbsub#1}}
\def\bbsmall#1{\hbox{\mybbsmall#1}}

\def\ZZ {\bb{Z}}
\def\ZZsub {\bbsub{Z}}

\def\RR {\bb{R}}
\def\RRslightlylarge {\bbslightlylarge{R}}
\def\RRlarge {\bblarge{R}}
\def\RRsub {\bbsub{R}}
\def\RRsmall {\bbsmall{R}}

\def\beq{\begin{equation}}
\def\eeq{\end{equation}}
\def\bdm{\begin{displaymath}}
\def\edm{\end{displaymath}}
\def\beqa{\begin{eqnarray}}
\def\eeqa{\end{eqnarray}}
\def\n{\nonumber\\}

\newcommand{\vs}[1]{\vspace{#1 mm}}

\begin{document}
\begin{titlepage}

\setcounter{page}{0}
\begin{flushright}
KEK Preprint 99-189 \\
hep-th/0003053
\end{flushright}

\vs{5}
\begin{center}
{\Large\bf Modular Invariant Critical Superstrings on 
Four-dimensional Minkowski Space $\times$ Two-dimensional 
Black Hole\\}
\vs{10}

{\large
Shun'ya Mizoguchi\footnote{shunya.mizoguchi@kek.jp}} \\
\vs{5}
{\it Institute of Particle and Nuclear Studies \\
High Energy Accelerator Research Organization (KEK) \\
Oho 1-1, Tsukuba, Ibaraki 305-0801, Japan} \\
\end{center}
\vs{10}
\centerline{{\bf{Abstract}}}
\vskip 3mm
Extending the seminal work of Bilal and Gervais, we construct 
a tachyon-free, modular invariant partition function for critical 
superstrings on four-dimensional Minkowski $\times$ two-dimensional 
black hole. This model may be thought of as an $SL(2,\RR)/U(1)$ 
version of Gepner models and corresponds to a conifold point on the 
moduli space of Calabi-Yau compactifications. We directly deal with 
$N=2$, $c=9$ unitary superconformal characters. Modular invariance 
is achieved by requiring the string to have a momentum along 
an extra noncompact direction, in agreement with the picture of 
singular CFTs advocated by Witten. The four-dimensional massless 
spectrum coincides with that of the tensionless strings, 
suggesting a possible dual description of type II strings on a 
conifold in terms of two intersecting NS5-branes. An interesting 
relation to $D=6$, $N=4$ gauged supergravity is also discussed.


\end{titlepage}

\baselineskip=18pt
\setcounter{footnote}{0}

\section{Introduction}

One of the issues of string theory preventing phenomenological 
applications is the moduli that we get upon compactification.
Since the critical dimension is $D=10$ for (worldsheet) 
$N=1$ superstrings, one has to compactify six of ten dimensions 
to get four-dimensional spacetime. Then there appear a number of 
massless scalars which characterize the size, the shape and other 
structures of the compactification space, although no so many 
(approximately) massless scalars are expected to be observed in 
reality. One must also break supersymmetries. It would therefore   
be interesting to explore any possibility to find a 
lower-dimensional string model with less supersymmetries 
which we could use as a starting point.

Recently, an interesting duality between strings on singular 
Calabi-Yau spaces and lower-dimensional non-gravitational 
(string) theories has been discussed 
\cite{Aharony:1998ub}-\cite{Pelc:2000hx}.  
In this paper, we will construct a modular invariant string partition 
function on four-dimensional Minkowski $\times$ two-dimensional black 
hole \cite{Witten:1991yr}. 
This model may be thought of as a noncompact $SL(2,\RR)/U(1)$ 
version of Gepner models \cite{Gepner} and corresponds 
\cite{Ghoshal:1995wm,Ooguri:1996wj} 
to a conifold point on the moduli space of Calabi-Yau 
compactifications of type II string theories. We directly deal with 
the characters of $N=2$, $c=9$ unitary superconformal field theory 
realized as a Kazama-Suzuki model \cite{Kazama:1989qp} based on 
$SL(2,\RR)/U(1)$ \cite{Dixon:1989cg} at level $k=3$. Unlike the 
conventional Gepner models, which use a tensor product of minimal 
models (realized as a compact coset $SU(2)/U(1)$ \cite{su(2)/u(1)}) 
as the internal CFT and is known \cite{Martinec:1989zu,Vafa:1989uu}
to describe regular Calabi-Yau 
compactifications, the necessary central charge 9 is supplied by 
a single $SL(2,\RR)/U(1)$ conformal field theory. Thus 
we do not need to take a tensor product. If we regard the coset 
conformal field theory as a gauged $SL(2,\RR)$ WZW model 
\cite{Schnitzer:1989qj,Witten:1992mk}, we end up with 
a six dimensional string theory on four-dimensional Minkowski 
space $\times$ two-dimensional black hole\footnote{
This is a {\it critical} string theory since the total central 
charge is 0 and the Liouville mode decouples. However, in the 
linear dilaton region which is far from the tip of the cigar, 
the coordinate field along the cigar looks like the Liouville 
field \cite{Witten:1991yr} which adjusts the total central to 0
if treated as a conformal field \cite{Distler:1989jt}. 
The idea of replacing the Liouville $\times$ $U(1)$ system in 
noncritical string theories with two-dimensional Euclidean black 
hole was proposed in \cite{Aharony:1998ub}.
}.

The main obstacle in constructing modular invariants using $N=2$, 
$c>3$ superconformal characters 
\cite{Dobrev:1987hq}-\cite{Kiritsis:1988rv} 
was its bad 
modular behavior (See \cite{Huitu:1993wm} for an earlier attempt.); 
for example, the nondegenerate NS character is given by 
\beqa
\mbox{Tr}q^{L_0}=q^{h+1/8}\frac{\vartheta_3(0|\tau)}{\eta^3(\tau)},
\eeqa
whose monomial factor $q^{h+1/8}$ makes the modular behavior 
awful. To overcome this problem, we use the following 
two ideas : First, we consider (a countable set of) infinitely 
many primary fields so that the sum of their $q^{h+1/8}$ 
factors form a certain theta function. Which theta function to choose 
has to be examined carefully, and will be determined later.
The modular property of the total partition function can be 
thus improved. Consistent CFTs with an infinite number of primary 
fields are known 
({\it e.g.} $c=1$ CFTs \cite{Yang:1987mf,Ginsparg:1988eb}) and  
not surprising. On the contrary, they are required for 
modular invariance \cite{Dixon:1989cg}.

There is still a problem even after the monomial of $q$ is replaced 
by a theta function; the number of eta and theta functions do not 
balance between the denominator and the numerator. It is a problem 
because in the modular $S$ transformation the $\sqrt{\tau}$ factors 
do not cancel. The situation is similar for a free scalar 
partition function, in which, however, the $|\tau|$ factor from 
the left and right eta functions is compensated by the modular 
transformation of the zeromode integral. Thus our second proposal 
is to assume that the ``internal'' $N=2$ superconformal system 
constructed from the $SL(2,\RR)/U(1)$ coset has a degree of freedom 
of the center-of-mass motion along a certain direction, which has to 
be integrated over in the partition function. This assumption just 
agrees with the picture of CFTs on singular Calabi-Yau spaces 
advocated by Witten \cite{Witten:1995zh}. With these two ideas 
we construct a modular invariant, and it turns out that 
the integration over the continuous set of ensembles above is 
nothing but the ``Liouville-momentum'' integration along the cigar
(See \cite{Dijkgraaf:1992ba} for an earlier analysis of strings on 
two-dimensional black hole.).

Another important question in constructing a model is how to restore 
the spacetime supersymmetry. In type II string theories the key role 
was played by Jacobi's abstruse identity. Some similar useful theta 
identities were found by Bilal and Gervais 
\cite{Bilal:1987uh,Bilal:1987ia} long time ago and were used to 
construct interesting noncritical superstring models. In fact, 
our model is closely related to their six-dimensional ($d=5$) model, 
in particular contains the latter as a subsector, although the 
interpretation is somewhat different.

The remainder of this paper is organized as follows. In sect.~2, 
we briefly review the construction of $N=2$ superconformal 
algebra based on the noncompact coset $SL(2,\RR)/U(1)$ and the 
geometrical meaning of the free fields used in the realization.
In sect.~3, we use theta identities to construct a modular invariant 
partition function, clarify the $N=2$ character content and 
summarize the features of our model. In sect.~4 we study the lowest 
mass spectra. Finally, we present our conclusions and suggestions 
for further work in sect.~5.

\section{$SL(2,\RRlarge)/U(1)$ Kazama-Suzuki model}
\subsection{Free-field realizations}
Let us begin with a review of the $SL(2,\RR)/U(1)$ Kazama-Suzuki 
model \cite{Dixon:1989cg}. We use the following free-field realization 
of the $SL(2,\RR)$ current algebra:
\beqa
J^3(z)&=&i\sqrt{~\frac{k}2~}\partial\phi,\n
J^\pm(z)&=&i\left(\sqrt{~\frac{k}2~}\partial\theta
                  \pm i\sqrt{\frac{k-2}2}\partial\rho\right)
         \exp\left(\pm i\sqrt{~\frac2{k}~}(\theta-\phi)
             \right).
\label{free-field_realization}
\eeqa
The OPEs of the free scalars are  
$\rho(z)\rho(0)\sim \theta(z)\theta(0)\sim -\log z$ but
$\phi(z)\phi(0)\sim +\log z$. This realization may be obtained by 
bosonizing \cite{Friedan:1986ge} the $\beta$-$\gamma$ system of the 
Wakimoto realization \cite{Wakimoto:1986gf}
followed by a redefinition of the scalars. The same realization was 
used in \cite{Satoh:1998xe}.
The energy-momentum tensor is 
\beqa
T_{SL(2,\RRsub)}(z)&=&
-\frac12(\partial\rho)^2 +\frac1{\sqrt{2(k-2)}}\partial^2\rho
-\frac12(\partial\theta)^2 +\frac12(\partial\phi)^2.
\eeqa
The central charge is $c_{SL(2,\RRsub)}=3k/(k-2)$.

The $SL(2,\RR)$ parafermions are defined by removing the exponentials 
of $\phi$ from $J^{\pm}$:
\beqa
\psi^\pm(z)&=&
i\left(\sqrt{~\frac12~}\partial\theta
       \pm i\sqrt{\frac{k-2}{2k}}\partial\rho\right)
         \exp\left(\pm i\sqrt{~\frac2{k}~}\theta
             \right),
\eeqa
where the parafermion fields $\psi^\pm$ are $\psi_1$ and  
$\psi_1^\dagger$ in the usual notation. (See also 
\cite{Griffin:1991fg} for a realization of the $SL(2,\RR)$ 
parafermion algebra.)

The energy-momentum tensor 
of the parafermion theory is 
\beqa
T_{SL(2,\RRsub)/U(1)}(z)&=&
T_{SL(2,\RRsub)}(z)
-\left(+\frac12(\partial\phi)^2\right).
\eeqa

The $N=2$ superconformal algebra can be obtained by adding 
back another free boson $\varphi(z)$ with the OPE 
$\varphi(z)\varphi(0)\sim-\log z$. The currents are given by
\beqa
T_{N=2}(z)&=&T_{SL(2,\RRsub)/U(1)}(z)-\frac12(\partial\varphi)^2,\n
T_F^{\pm}(z)&=&\sqrt{\frac{2k}{k-2}}\psi^\pm
\exp\left(
\pm i\sqrt{\frac{k-2}{k}}\varphi
\right),\\
J_{N=2}(z)&=&i\sqrt{\frac{k}{k-2}}\partial\varphi.\nonumber
\eeqa
The central charge is $c_{N=2}=3k/(k-2)$ again. $k$ will be set to 
3 later. 

\subsection{Representations of classical SL(2,\RRslightlylarge)}
A unitary $SL(2,\RR)/U(1)$ coset module is constructed by forbidding 
the $J^3$ excitation in a unitary highest (or rather lowest) weight 
module of the $SL(2,\RR)$ current algebra. Thus the lowest $L_0$ level 
states form a unitary representation of (classical) $SL(2,\RR)$.  
(See {\it e.g.} \cite{VK} for the representations of the classical 
$SL(2,\RR)$ Lie algebra. See also \cite{Dijkgraaf:1992ba}.)
The latter can be realized as the differential operators acting 
on functions on $S^1$:
\footnote{These equations correct the inconsistency in the sign 
convention in ref.~\cite{VK}.}
\beqa
H^3&=&\epsilon+i\frac{d}{dx},\n
H^{\pm}&=&ie^{\mp ix}\frac{d}{dx}+(\epsilon\pm l)e^{\mp ix},\\
C&=&\frac12(H^+H^- +H^-H^+)-(H^3)^2.\nonumber
\eeqa
$\{e^{-imx}~|~m\in\ZZ\}$ form a complete set and 
\beqa
H^3 e^{-imx}&=&(m+\epsilon) e^{-imx},\n
H^\pm e^{-imx}&=&(m+\epsilon\pm l) e^{-i(m\pm 1)x},\\
C e^{-imx}&=&-l(l-1) e^{-imx}.\nonumber
\eeqa
Thus the representations of $SL(2,\RR)$ are labeled by $(l,\epsilon)$. 
The corresponding equations in the $SL(2,\RR)$ current algebra module 
are 
\beqa
J^3_0 |m+\epsilon>&=&(m+\epsilon)|m+\epsilon>,\n
J^\pm_0 |m+\epsilon>&=&(m+\epsilon\pm l)|m+\epsilon\pm 1>,\\
\mbox{\boldmath $J$}|m+\epsilon>&=&-l(l-1)|m+\epsilon>\nonumber
\eeqa
with $\mbox{\boldmath $J$}=1/2(J_0^+ J_0^- +J_0^- J_0^+)-(J_0^3)^2$.

Unitary representations $(l,\epsilon)$ of (classical) $SL(2,\RR)$ 
are known to be classified into the following four cases:\footnote{
We do not need to consider the universal covering of 
$SL(2,\RRsmall)$ in our $c_{N=2}=9$ case. Then $\epsilon$ takes 
discrete values.}

\begin{itemize}

\item[{\it i.}]{{\it Principal unitary series}: 
$(+\frac12+ip,\epsilon)$; $p\in\RR$, $\epsilon\in\{0, \frac12\}$.}

\item[{\it ii.}]{{\it Complementary series}: 
$(l,0)$; $0<l<1$.}

\item[{\it iii$+$.}]{{\it Discrete series} ${\cal D}_n^+$: 
$(n+\epsilon,\epsilon)$; $\epsilon\in\{0, \frac12\}$,
$n\in\ZZ_{>0}$ if $\epsilon=0$, $n\in\ZZ_{\geq 0}$ if $\epsilon=\frac12$.} 

\item[{\it iii$-$.}]{{\it Discrete series} ${\cal D}_n^-$: 
$(n-\epsilon,\epsilon)$; $n\in\ZZ_{\geq 0}$, $\epsilon\in\{0, \frac12\}$.} 

\item[{\it iv.}]{{\it Trivial representation}.} 

\end{itemize}

In the cases {\it i}, {\it ii}, the whole module consists of the 
states $|m+\epsilon>$, $m\in\ZZ$. There are neither highest weight 
states nor lowest-weight states.
In the cases {\it iii$\pm$}, the module spanned by $|m+\epsilon>$, 
$m\in\ZZ$ turns reducible due to the appearance of null states. 
In this case the irreducible submodules ${\cal D}_{l\mp\epsilon}^\pm$ 
are unitary.  ${\cal D}_{l-\epsilon}^+$ 
(${\cal D}_{l+\epsilon}^-$ ) has a lowest- 
(highest-)weight state $|l>$ ($|-l>$). 
The quotient module divided by 
${\cal D}_{l-\epsilon}^+\oplus{\cal D}_{l+\epsilon}^-$  
is finite, and non-unitary if $l\neq\frac12,1$. If $l=\frac12$ the
quotient module is empty; if $l=1$ the quotient module consists of 
a single state, and hence corresponds to the trivial representation. 
One can also construct unitary representations by starting from 
negative $n$, but they only give equivalent representations. 
(${\cal D}_{0}^+$ is equivalent to ${\cal D}_{1}^+$ if $\epsilon=0$.)  
This is manifest in the symmetry of the Casimir $l\leftrightarrow-l+1$.
Finally, the trivial representation maps any element of the Lie 
algebra to 0.

The vertex operator of the $SL(2,\RR)$ current algebra at level $k$ 
corresponding to the state $|m+\epsilon>$ in the $(l,\epsilon)$ 
representation is
\beqa
|m+\epsilon>\sim e^{+\sqrt{\frac2{k-2}}l\rho
                     +i\sqrt{\frac2k}(m+\epsilon)(\theta-\phi)}.
\eeqa
It has 
\beqa
L_0^{SL(2,\RRsub)}=-\frac{l^2-l}{k-2}, 
~~~J^3_0=m+\epsilon.
\eeqa
The corresponding $N=2$ vertex operator is
\beqa
e^{+\sqrt{\frac2{k-2}}l\rho
                     +i\sqrt{\frac2k}(m+\epsilon)\theta
		     +i\frac{2(m+\epsilon)}{\sqrt{k(k-2)}}\varphi}.
\label{vertex_op}
\eeqa
It has 
\beqa
h=L_0^{N=2}=\frac{-(l^2-l)+(m+\epsilon)^2}{k-2},
~~~Q=J_0^{N=2}=-\frac{2(m+\epsilon)}{k-2}.
\eeqa
\vskip 2ex

\subsection
{Interpretation ---
the $SL(2,\RRslightlylarge)$ WZW model} 

One of the nice features of the realization (\ref{free-field_realization})
is its clear geometrical meanings. To see this, let us write out 
the $SL(2,\RR)$ WZW action
\beqa
S_{\mbox{\scriptsize WZW}}
&=&\frac{k}{8\pi}\int d^2x \mbox{Tr} \partial_\mu g \partial^\mu g^{-1}
   +k\Gamma(g),\\
\Gamma(g)&=&\frac1{12\pi}\int d^3x 
            \epsilon^{\overline{\mu}\overline{\nu}\overline{\rho}}
            \mbox{Tr}
            \overline{g}^{-1}\partial_{\overline{\mu}}\overline{g}
            ~\overline{g}^{-1}\partial_{\overline{\nu}}\overline{g}
            ~\overline{g}^{-1}\partial_{\overline{\rho}}\overline{g}
\eeqa
using the parameterization
\beqa
g&=&\left[\begin{array}{rl}
u+w&v+y\\-v+y&u-w
\end{array}\right],
\eeqa
\[
\begin{array}{ll}
u~=~\cosh\mbox{\boldmath $\rho$}
\cos\mbox{\boldmath $t$},~~~~~&
v~=~\cosh\mbox{\boldmath $\rho$}
\sin\mbox{\boldmath $t$},\\
w~=~\sinh\mbox{\boldmath $\rho$}
\cos\mbox{\boldmath $\tilde{\theta}$},~~~~~&
y~=~\sinh\mbox{\boldmath $\rho$}
\sin\mbox{\boldmath $\tilde{\theta}$}.
\end{array}
\]
(The signature of the worldsheet is 
$\eta^{\mu\nu}=\mbox{diag}[-1,+1]$.)
The result is 
\beqa
S_{\mbox{\scriptsize WZW}}
&=&-\frac{k}{8\pi}\int d^2x\left[\sqrt{-h}h^{\mu\nu}
(\partial_\mu\mbox{\boldmath $\rho$}\partial_\nu\mbox{\boldmath $\rho$}
-\cosh^2\!\!\mbox{\boldmath $\rho$}
\partial_\mu\mbox{\boldmath $t$}\partial_\nu\mbox{\boldmath $t$}
+\sinh^2\!\!\mbox{\boldmath $\rho$}
\partial_\mu\mbox{\boldmath $\tilde{\theta}$}
\partial_\nu\mbox{\boldmath $\tilde{\theta}$})\right.\n
&&\left.~~~~~~~~~~~~~~~~~~~~
-2\epsilon^{\mu\nu}\sinh^2\!\!\mbox{\boldmath $\rho$}
\partial_\mu\mbox{\boldmath $t$}
\partial_\nu\mbox{\boldmath $\tilde{\theta}$}
\right]\n
&=&\frac{k}{4\pi}\int d^2x\left[
-\partial_+\mbox{\boldmath $\rho$}\partial_-\mbox{\boldmath $\rho$}
+\partial_+\mbox{\boldmath $t$}\partial_-\mbox{\boldmath $t$}
+\sinh^2\!\!\mbox{\boldmath $\rho$}
\partial_+(\mbox{\boldmath $t$}
-\mbox{\boldmath $\tilde{\theta}$})
\partial_-(\mbox{\boldmath $t$}
+\mbox{\boldmath $\tilde{\theta}$})\right].
\eeqa
The last line is the expression on the flat worldsheet with
$x^\pm=(x^1\pm x^0)/\sqrt{2}$. 
$SL(2,\RR)$ invariant metric is given by 
\beqa
ds^2&=&d\mbox{\boldmath $\rho$}^2
-\cosh^2\mbox{\boldmath $\rho$} d\mbox{\boldmath $t$}^2
+\sinh^2\mbox{\boldmath $\rho$} d\mbox{\boldmath $\tilde{\theta}$}^2
\eeqa
in this parameterization. 
We now dualize 
\mbox{\boldmath $\tilde{\theta}$} by adding  
\beqa
-\frac{k}{4\pi}\int d^2x \epsilon^{\mu\nu}
\partial_\mu\mbox{\boldmath $\theta$}
\partial_\nu\mbox{\boldmath $\tilde{\theta}$}
\eeqa
to $S_{\mbox{\scriptsize WZW}}$. The dual action reads 
\cite{Buscher,Horowitz:1993jc}
\beqa
S^{\mbox{\scriptsize dual}}_{\mbox{\scriptsize WZW}}
&=&-\frac{k}{8\pi}\int d^2x\sqrt{-h}\left[
(\partial_\mu\mbox{\boldmath $\rho$})^2
-(\partial_\mu(\mbox{\boldmath $t$}+\mbox{\boldmath $\theta$}))^2
+\tanh^2\!\!\mbox{\boldmath $\rho$}
(\partial_\mu\mbox{\boldmath $\theta$})^2\right]
\eeqa
with the dilaton field
\beqa
\Phi=-\log\cosh\mbox{\boldmath $\rho$}+\mbox{const.}
\eeqa
Thus the target space of the dual sigma model is $\RR\times$
two-dimensional black hole. The cigar geometry may be obtained 
by simply dropping the ``time'' coordinate 
\mbox{\boldmath $t$}+\mbox{\boldmath $\theta$}.
In the region $\mbox{\boldmath $\rho$}\rightarrow\infty$, 
$\mbox{\boldmath $\rho$}$, $\mbox{\boldmath $\theta$}$
and $\mbox{\boldmath $t$}+\mbox{\boldmath $\theta$}$  
correspond to the free fields $\rho$, $\theta$ and $\phi$, 
respectively. Clearly, $\rho$ plays the role of the Liouville 
field. 

\section{Modular Invariant Partition Function}

Having reviewed the $SL(2,\RR)/U(1)$ coset construction, we will 
now construct a modular invariant partition function. In type II 
string theories, the key equation was 
Jacobi's abstruse identity: 
\beqa
\vartheta_3^4(0|\tau)-\vartheta_4^4(0|\tau)
-\vartheta_2^4(0|\tau)=0. \label{abstruse} 
\eeqa
Are there any such nice theta identities for us, too? 
In fact, we may find one in the works on noncritical 
string theory done by Bilal and Gervais long time ago: 
\beqa
\Lambda_1(\tau)&\equiv&\Theta_{1,1}(\tau,0)
\left(\vartheta_3^2(0|\tau)+\vartheta_4^2(0|\tau)\right)
-\Theta_{0,1}(\tau,0)
\;\vartheta_2^2(0|\tau)~=~0,
\label{theta_id}
\eeqa
where 
\beqa
\Theta_{m,1}(\tau,\nu)&=&\sum_{n\in\ZZsub}q^{(n+m/2)^2}z^{2(n+m/2)}
~~~(m=0,1),
\eeqa
\[q=\exp(2\pi i \tau),~~z=\exp(2\pi i \nu)\]
are the level-1 $SU(2)$ theta functions. 
We also use another identity :
\beqa
\Lambda_2(\tau)&\equiv&\Theta_{0,1}(\tau,0)
\left(\vartheta_3^2(0|\tau)-\vartheta_4^2(0|\tau)\right)
-\Theta_{1,1}(\tau,0)
\;\vartheta_2^2(0|\tau)~=~0,
\label{theta_id'}
\eeqa
which is nothing but the modular $S$ transform of (\ref{theta_id}).
The determinant of the coefficient matrix of $\Theta_{m,1}(\tau,0)$ 
in (\ref{theta_id})(\ref{theta_id'})
consistently vanishes (for they are nonzero functions) due to 
(\ref{abstruse}).
Their modular properties are  
\beqa
\Lambda_1(\tau+1)&=&i\Lambda_1(\tau),\n
\Lambda_2(\tau+1)&=&-\Lambda_2(\tau),
\eeqa
and
\beqa
\Lambda_1(\tau)&=&\frac{\exp(3\pi i/4)}{\sqrt{2}\tau^{3/2}}
\left(-\Lambda_1(-1/\tau)+\Lambda_2(-1/\tau)\right),\n
\Lambda_2(\tau)&=&\frac{\exp(3\pi i/4)}{\sqrt{2}\tau^{3/2}}
\left(+\Lambda_1(-1/\tau)+\Lambda_2(-1/\tau)\right).
\eeqa
Thus 
\beqa
\left|\Lambda_1(\tau)/\eta^3(\tau)\right|^2
+\left|\Lambda_2(\tau)/\eta^3(\tau)\right|^2
\label{modular_invariant}
\eeqa
is modular invariant. 

Having found a building block, we now construct a partition 
function. We set $k=3$ from now on.
For the NS sector, we collect the unitary representations
\cite{N=2Kac,Boucher:1986bh}
\beqa
h=\frac{Q^2+1}4+p^2,~~~Q\in\ZZ \label{NSrepresentations}
\eeqa 
and integrate over $p\in\RR$ with the same weight (Figure.1). 
Except when $Q$ is odd and $p=0$, they all have non-degenerate 
characters: 
\beqa
\mbox{Tr}_{\rm NS}q^{L_0^{N=2}}z^{J_0^{N=2}}={\rm ch}_{\rm NS}(h,Q)
=q^{p^2+\frac{Q^2+1}4}z^Qf^{\rm NS}(\tau,z),
\eeqa
where 
\beqa
f^{\rm NS}(\tau,z)&\equiv&\prod_{n=1}^\infty 
                   \frac{(1+zq^{n-1/2})(1+z^{-1}q^{n-1/2})}
                        {(1-q^n)^2}
~~~\left(
=q^{1/8}\frac{\vartheta_3(\nu|\tau)}{(\eta(\tau))^3}
\right).
\eeqa
Note that this class of representations precisely correspond to 
the ones made out of the principal unitary series only. 

When 
\beqa
Q=Q_m=-(2m+1),~~~
h=h_m=\frac{Q_m^2+1}4=m^2+m+\frac12
~~~(m\in\ZZ)
\eeqa 
(that is, $p=0$), 
the representation reaches the boundary of the unitarity region, 
where its irreducible subspace becomes smaller.    
The non-degenerate character then decomposes \cite{Eguchi:1988af} 
into a sum of two degenerate characters of A$_3$ type: 
\beqa
{\rm ch}_{\rm NS}(h_m, Q_m)
={\rm ch}_{{\rm A}_3{\rm deg}}(h_m, Q_m; m)
+{\rm ch}_{{\rm A}_3{\rm deg}}(h_m+|m+1/2|, 
			       Q_m+\mbox{sign}(Q_m); m),
\label{character_Higgs_mechanism}
\eeqa
where
\beqa
{\rm ch}_{{\rm A}_3{\rm deg}}(h, Q; r)
=q^{h}z^{Q} \frac{f^{\rm NS}(\tau,z)}{1+q^{|r+1/2|}z^{{\rm sign}(Q)}}.
\eeqa
The first term of (\ref{character_Higgs_mechanism}) 
is the irreducible character of the representation $(h_m, Q_m)$, 
while the second is that of the adjacent integer $Q$ degenerate 
representation on the same boundary line (Figure.1). 
These representations are also summed over. As a result, the 
points $(h_m,Q_m)$ may also be thought of as if they were generic 
(non-degenerate) ones. Such degenerate representations on the 
boundary lines of the unitarity region are made out of the discrete 
series of $SL(2,\RR)$.

For the R sector, we similarly consider the following set of 
representations:   
\beqa
h=\frac{Q^2}4+p^2,~~~Q\in\ZZ.\label{Rrepresentations}
\eeqa
In this case, unless $Q$ is even and $p=0$, the characters are 
given by the generic ones: 
\beqa
\mbox{Tr}_{\rm R}q^{L_0^{N=2}}z^{J_0^{N=2}}={\rm ch}_{\rm R}(h,Q)
=q^{p^2+\frac{Q^2}4}z^Qf^{\rm R}(\tau,z),
\eeqa
where
\beqa
f^{\rm R}(\tau,z)&\equiv&(z^{1/2}+z^{-1/2})
                   \prod_{n=1}^\infty 
                   \frac{(1+zq^n)(1+z^{-1}q^n)}
                        {(1-q^n)^2}
~\left(
=\frac{\vartheta_2(\nu|\tau)}{(\eta(\tau))^3}
\right).
\eeqa
We do not shift the $U(1)$ charge by $\pm 1/2$ 
in the definition of the Ramond character; if 
the representation has two degenerate lowest $L_0$  
states (which is the generic ($h\neq 3/8$) case), 
$Q$ represents the {\it mean} value of $U(1)$ 
charges of the two states. If $h=3/8$, the lowest 
$L_0$ state is unique because the other becomes null. 
Then the $U(1)$ charge of the lowest $L_0$ state 
is $Q-1/2$ (P$^+$ module).

If $Q$ is even and $p=0$, the non-degenerate character again can be 
written as a sum of two irreducible degenerate characters 
of P$_3^+$ type: 
\beqa
{\rm ch}_{\rm R}(h_m, Q_m)
={\rm ch}_{{\rm P}_3^+{\rm deg}}(h_m, Q_m; m)
+{\rm ch}_{{\rm P}_3^+{\rm deg}}
	(h_m+|m|, Q_m+\mbox{sign}(Q_m+1/2); m),
\label{Rcharacter_Higgs_mechanism}
\eeqa
where
\beqa
Q=Q_m=-2m,~~~
h=h_m=\frac{Q_m^2}4=m^2
~~~(m\in\ZZ)
\eeqa 
and
\beqa
{\rm ch}_{{\rm P}_3^+{\rm deg}}(h, Q; r)
=q^{h}z^{Q} \frac{f^{\rm R}(\tau,z)}{1+q^{|r|}z^{{\rm sign}(Q+1/2)}}
\eeqa
(Figure.2).
Again, taking into account the extra degenerate representations 
(the second term of (\ref{Rcharacter_Higgs_mechanism})), 
the points $(h_m, Q_m)$ can be thought of as generic.  

The two sets of representations (\ref{NSrepresentations}) and 
(\ref{Rrepresentations}), as well as the extra degenerate 
representations added to them, transform into each other by a
spectral flow.

Which GSO projection should we take? We wish to construct a 
partition function in which the fermion theta in the 
four-dimensional Minkowski + ghost sector and the characters 
of the internal $N=2$ sector are combined into the form like 
(\ref{modular_invariant}). Thus we need to have the following 
GSO projection: 
Let $F$ 
($\overline{F}$) be the right (left) fermion number of the 
four-dimensional Minkowski + ghost sector, $Q$ ($\overline{Q}$) 
the right (left) $N=2$ $U(1)$ charge and $\epsilon$ 
($\overline{\epsilon}$) the parameter which distinguishes 
the parity of the $U(1)$ charge of the right (left) $N=2$ 
ground state (See eq.~(\ref{vertex_op}).). 
Then 
for the NS sector, we only keep the states with 
both $F+Q$ and $\overline{F}+\overline{Q}$ odd, 
{\it and} $2\epsilon+2\overline{\epsilon}$ even. 
In other words, we keep odd fermion excited states if 
$\epsilon=\overline{\epsilon}=0$, 
while we do even states if $\epsilon=\overline{\epsilon}=1/2$. 
The two parameters $m$ and $\epsilon$ labeling the $U(1)$ 
charge are separately GSO projected. 
On the other hand, for the R sector, the states with the same $N=2$ 
vacuum $U(1)$ parity ({\it i.e.} $\epsilon=\overline{\epsilon}$) 
are similarly paired, but the left-right chirality may or may not 
be the same. If the chirality is the same, we get a IIB-like
model, while if it is opposite, we get a IIA-like model.
In addition, the left-right diagonal $p=\overline{p}$ are required
for modular invariance for both sectors.

With this GSO projection, the total partition function reads
\beqa
Z(\tau)=\int\frac{d\tau d\overline{\tau}}{\mbox{Im}\tau}
	(\mbox{Im}\tau)^{-2}\left|\eta(\tau)\right|^{-4}
(\mbox{Im}\tau)^{-\frac12}\left|\eta(\tau)\right|^{-2}
\left[\left|\Lambda_1(\tau)/\eta^3(\tau)\right|^2
+\left|\Lambda_2(\tau)/\eta^3(\tau)\right|^2\right],
\label{partition_function}
\eeqa
where the factor $(\mbox{Im}\tau)^{-\frac12}$ comes 
from the diagonal ``Liouville-momentum'' $p$ integration, and 
$\left|\eta(\tau)\right|^{-2}$ from the transverse fermions.
The transverse fermion theta has already been taken into account 
in the last factor and GSO projected with the $N=2$ theta together. 
This is the main result of this paper. 

We will now list some of the notable features of our partition 
function $Z(\tau)$ (\ref{partition_function}): 

\begin{itemize}

\item[{\it 1.}]{{\it It is modular invariant.} 
Modular invariance has been achieved by integrating the 
``Liouville momenta'' $p$. This may be understood as a summation 
over the radial momenta on the cigar. Indeed, the principal unitary 
series is the only class of representations that corresponds to 
an $N=2$ vertex operator with {\it real} $\rho$ momentum (apart 
from the imaginary background charge $-i/\sqrt{2}$). Consequently, 
spacetime, which was supposed to be four-dimensional, turns 
five-dimensional effectively; any ``particle'' in the 
four-dimensional world has a continuous spectrum.  
This agrees with the picture of singular CFTs advocated in 
ref.~\cite{Witten:1995zh}.
}

\item[{\it 2.}]{{\it It is unitary and tachyon free.} 
}

\item[{\it 3.}]{{\it It is spacetime supersymmetric.} $Z(\tau)$ is zero
in reality. Since the vanishing $\Lambda_{1,2}(\tau)$ are a consequence 
of the ordinary spectral flow, the spacetime supercharge must be 
given by the usual one using the bosonized ghost, the fermion-number 
current and the $N=2$, $U(1)$ current. This is in contrast to the one 
in ref.~\cite{Kutasov:1990ua} containing a contribution from the 
``longitudinal'' boson.
}

\item[{\it 4.}]{{\it It has a graviton.} 
The tensor product of the NS transverse fermion excitations yields 
a graviton, a dilaton and an anti-symmetric two-form field. 
They survive the GSO projection. 
This is the most significant difference between Bilal-Gervais's model 
\cite{Bilal:1987uh,Bilal:1987ia} and ours; the graviton comes from 
the $F+Q-2\epsilon$ odd sector ($\Lambda_2(\tau)$), which is missing 
in the former. Those fields are massive in the sense that they have 
$L_0=\overline{L_0}=1/4$ even when $p=\overline{p}=0$. }

\item[{\it 5.}]{{\it It contains bound states in the spectrum.}
As we have shown in (\ref{character_Higgs_mechanism}) and 
(\ref{Rcharacter_Higgs_mechanism}), the partition function has a 
contribution from the representations made out of the discrete series 
of $SL(2,\RR)$. They do not have a momentum along the cigar and are 
regarded as the bound states \cite{Dijkgraaf:1992ba}.}

\end{itemize}

\section{Mass Spectra}
Let us now discuss more in detail the lightest mass spectra of our 
model. In the last section, we have seen that any four-dimensional 
particle exhibits a continuous mass spectrum, which is attributed to  
its momentum along the cigar. Therefore we consider the value of 
total $L_0$ of particles ``at rest'' along the cigar 
({\it i.e.} $p=0$). This is equivalent to studying masses in five 
dimensions.

Since the transverse fermion theta and the $N=2$ fermion 
theta are GSO projected together and enter in the partition function 
symmetrically, 
$SO(4)$ (acting on four real fermions) plays an analogous role to the 
transverse rotational group (or the little group for massive states) 
in six dimensions, although our model does not have six-dimensional 
Poincare invariance. The four-dimensional field content can be 
conveniently obtained by a dimensional reduction (since the $N=2$ 
fermions carry no spacetime indices).

We first consider the states coming from $|\Lambda_1(\tau)|^2$. The 
lightest NS-NS fields are four scalars. They are of course common 
in both types of GSO projection in the R sector. On the other hand, the 
doubly-degenerate lowest $L_0$ states in the R sector cannot be a spinor 
of $SO(4)$ because a pseudo-real spinor needs four components. 
Thus it can only be a nonchiral Majorana $SO(2)$ spinor. 
Then the lightest R-R fields are a four-dimensional vector and two 
scalars in either projection. (The IIB- and IIA-like projections 
yield the same R-R fields here because either of the eigenspaces 
of the $SO(4)$ chirality operator decompose into a direct sum of 
$+$ and $-$ $SO(2)$ chirality eigenspaces.) They are massless 
($L_0=\overline{L_0}=0$). Including fermions, they form 
a four-dimensional $N=2$ $U(1)$ vector multiplet
+ a hypermultiplet. They are the same field content as a single 
$N=4$, $U(1)$ vector multiplet has, although we have 
only eight spacetime supersymmetries from the standard supercharge 
construction. It is interesting that those fields are formally 
obtained by a dimensional reduction of a six-dimensional (2,0) 
tensor multiplet or a (1,1) vector multiplet on a flat background.
They are the lightest fields of Bilal-Gervais's closed string 
model \cite{Bilal:1987ia}.

The lightest NS-NS bosons from $|\Lambda_2(\tau)|^2$ are a graviton, 
a dilaton and a 2-form (self-dual + anti-self-dual) as we have seen 
in the previous section. In the R sector, we have now twice as many 
states as those in $|\Lambda_1(\tau)|^2$ at the lowest level, and 
hence may regard them as the components of a single $SO(4)$ Weyl 
spinor. Then in the R-R sector, the IIB-like projection yields four 
anti-self-dual 2-forms and four scalars, while the IIA-like projection  
gives four vectors. They are the bosonic fields of a six-dimensional 
$N=2$ (``${\cal N}=1$'') graviton + a self-dual tensor 
+ four anti-self-dual tensor multiplets (IIB-like), and a graviton + 
a self-dual tensor + four vector multiplets (IIA-like), respectively. 
Again, they combine into a (2,0) graviton + a tensor multiplets in 
the former case, and a single (1,1) graviton multiplet in the latter. 
The four-dimensional field contents are obtained by the dimensional 
reduction of those fields. They have $L_0=\overline{L_0}=1/4$.

\section{Conclusion}

In this paper we have constructed a modular invariant partition 
function of superstrings on four-dimensional Minkowski space 
$\times$ two-dimensional black hole using the $N=2$, $c=9$ 
superconformal characters. Our model may be thought of as a 
modular invariant extension of Bilal-Gervais's $d=5$ noncritical 
string model and describes type II strings on a conifold. 
It is unitary, tachyon free and has a continuous spectrum in 
the four-dimensional sense.

In the $\alpha'\rightarrow 0$ limit the cigar becomes 
very thin and can be replaced by a thin cylinder 
$\RR\times S^1$ since one would need very high energy to see the 
effect of the sharp tip. In this 
case one is left with a four-dimensional $N=2$ non-gravitational 
theory with a vector multiplet and a hypermultiplet. This will give  
an example of holography proposed in \cite{Aharony:1998ub,Giveon:1999zm}, 
and the $SL(2,\RR)$ coset in our model will provide a regularization 
of the strong coupling singularity of the linear-dilaton vacuum.

It is also interesting that the four-dimensional massless spectrum 
(with $p=0$) of our model coincides with that of the tensionless 
strings which arise on the four-dimensional intersection of two 
M5-branes \cite{Hanany:1996hq}. To understand its implication we 
recall that the (0,4) tensionless strings arise in type IIB theory 
on K3 when K3 gets an ADE singularity \cite{Witten:1995zh}, while  
IIB on such a singular K3 is known \cite{Ooguri:1996wj} to be T-dual 
to a system of type IIA 5-branes \cite{Callan:1991at}. Thus, in 
view of this, the coincidence of the spectra may suggest that type 
II strings on a Calabi-Yau threefold with a conifold singularity have 
a dual description in terms of two intersecting NS5-branes.

On the other hand, $\alpha'\rightarrow \infty$ means that the area 
near the tip of the cigar is zoomed in on, and the whole target 
space looks like a six-dimensional Minkowski space. In this case all 
the towers of particles affect the low-energy physics, of which we 
cannot expect any local quantum field theory descriptions.

In the middle region of $\alpha'$ in between, the lowest 
level fields do not completely decouple but interact with some 
other light fields. In the previous section, we have seen that 
the first two lightest fields in the IIA-like GSO projection are   
the dimensional reduction of a six-dimensional (1,1) vector multiplet 
and a graviton multiplet. Remarkably, they are the field content of 
$D=6$, $N=4$ ($=(1,1)$) gauged supergravity 
\cite{Giani:1984dw,Romans:1986tw}!  
Although the graviton multiplet has $L_0=\overline{L_0}=1/4$, 
perhaps this is linked to the well-known subtlety in defining 
masslessness of a particle in a curved background (See {\it e.g.} 
\cite{vanNieuwenhuizen:1985iz}.). Indeed, 
$L_0+\overline{L_0}$ corresponds to the Klein-Gordon operator in 
a flat space, but the correspondence becomes less clear in the 
Minkowski $\times$ cigar geometry. If the massless point is shifted 
by $1/4$, the graviton becomes massless but the lightest fields from 
$|\Lambda_1(\tau)|^2$ then have negative mass square. They can,
however, still remain stable if they are above the
Breitenlohner-Freedman bound \cite{Breitenlohner:1982bm}. 
It would be interesting to explore $D=6$, $N=4$ 
gauged supergravity on this background. The supergravity 
interpretation of the IIB-like projection model remains 
an open question.

Finally, some generalization of our construction to other 
singular Calabi-Yau spaces can be done when the corresponding 
analogue of Jacobi's abstruse identity is known. 
For example, an interesting theta identity was found in 
\cite{Kutasov:1991pv}, which seems to be related to a Calabi-Yau 
four-fold with a conifold singularity. Very recently, new theta 
identities corresponding to Calabi-Yau $n$-folds with an ADE 
singularity have been systematically obtained by Eguchi and 
Sugawara \cite{Eguchi:2000tc}.

\vskip 3ex \noindent
I wish to thank T.~Eguchi, A.~Fujii, T.~Kawai, Y.~Satoh, Y.~Yamada 
and S.-K.~Yang for valuable discussions. I also wish to thank 
K.~Fujikawa, Y.~Matsuo, Y.~Sugawara and all participants of the 
seminar given at Dept.~of Physics, University of Tokyo in 
Nov.~1999 for questions and comments, which were useful for 
completing the final version. I am also grateful for the 
hospitality of Dept.~of Math., Kobe University 
where a part of this work was done.

\newpage
\begin{figure}
\begin{center}
\epsfig{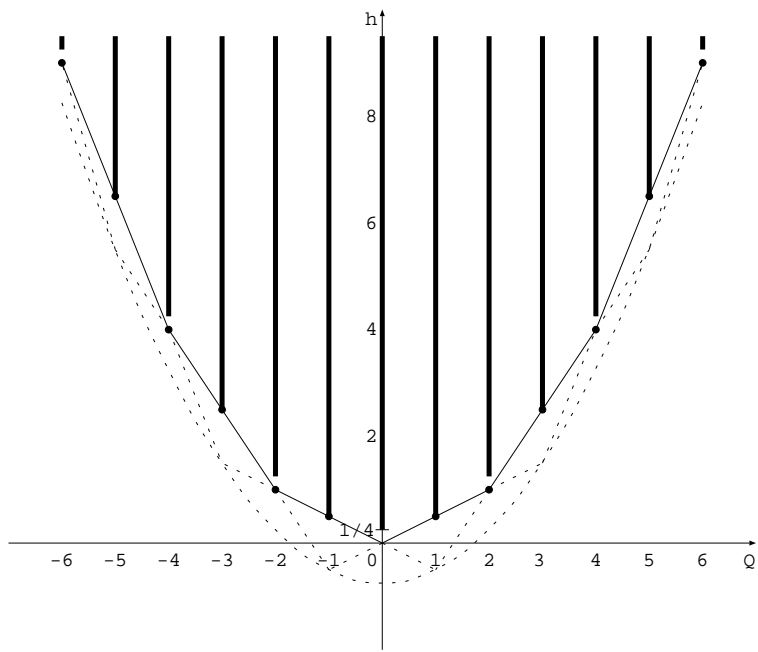}%
\end{center}
\caption{\label{figNS} 
The NS sector. The thick lines and dots show the $N=2$ unitary 
representations used as the internal CFT. The former correspond to 
the propagating modes along the cigar, while the latter are the 
bound states.}
\end{figure}
\begin{figure}
\begin{center}
\epsfig{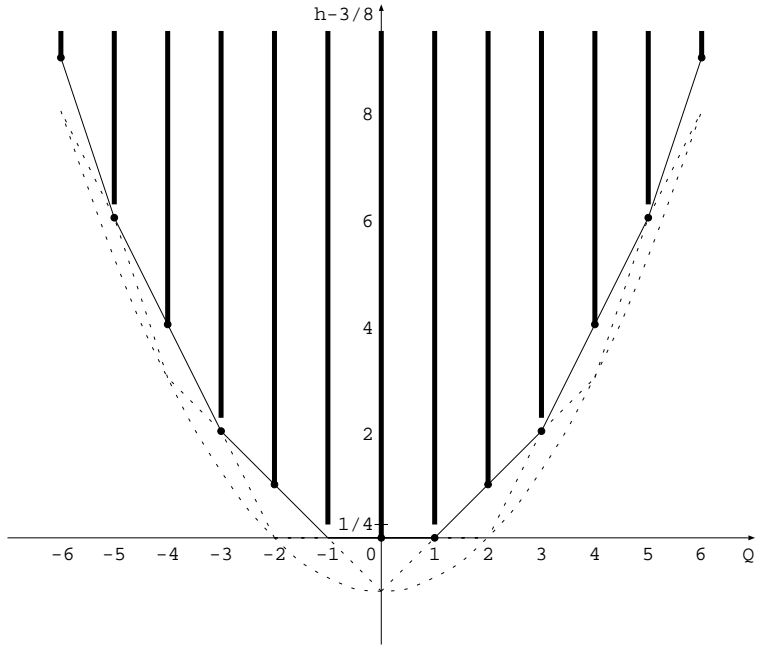}%
\end{center}
\caption{\label{figR} The R sector.  
$Q$ represents the mean value of $U(1)$ charges of two lowest $L_0$ 
states if $h\neq 3/8$. The $U(1)$ charge of 
the lowest weight state is $Q\pm 1/2$ depending on the convention 
(which of ${\rm P}^\pm$ representations is considered). If $h=3/8$, 
the lowest $L_0$ state is unique because the other becomes null. 
The (superficial) asymmetry at the bottom is due to the convention 
used here; we consider ${\rm P}^+$ representations so that the $U(1)$ 
charge of the lowest weight state is $Q-1/2$.}
\end{figure}

\end{document}